\documentclass[10pt,aps,prl,twocolumn,superscriptaddress,nofootinbib]{revtex4-1}
\usepackage{graphicx}
\usepackage{bigints}
\usepackage[normalem]{ulem}
\usepackage{cancel}
\usepackage{mathtools}
\usepackage{verbatim}
\usepackage{slashed}
\usepackage{titlesec}
\usepackage[usenames]{xcolor}
\usepackage{amsmath,amssymb}
\usepackage{mathrsfs}
\usepackage{amsthm}
\usepackage{hyperref}
\usepackage{appendix}
\usepackage{array}

\newcommand{\sss}{\scriptscriptstyle}

\newcommand{\p}{\partial}

\newcommand{\be}[1]{ \begin{equation}\label{#1} }
\newcommand{\ee}{\end{equation}}
\newcommand{\bea}[1]{\begin{eqnarray}\label{#1} }
\newcommand{\eea}{\end{eqnarray}}
\newcommand{\bes}{\begin{subequations}}
\newcommand{\ees}{\end{subequations}}

\newcommand{\<}{\langle}
\renewcommand{\>}{\rangle}

\renewcommand{\(}{\left(}
\renewcommand{\)}{\right)}

\newcommand{\lb}{\Big[}
\newcommand{\rb}{\Big]}

\begin{document}

\title{Towards Carrollian quantization: renormalization of Carrollian electrodynamics}
\author{Aditya Mehra}
\email{amehra@ed.ac.uk}
\affiliation{School of Mathematics and Maxwell Institute for Mathematical Sciences, University of Edinburgh, Peter Guthrie Tait Road, Edinburgh EH9 3FD, UK}

\author{Aditya Sharma}
\email{adityasharma.theory@gmail.com}
\affiliation{Department of Physics, BITS-Pilani, K K Birla Goa Campus, Zuarinagar, Goa-403726, INDIA}
\begin{abstract}
\footnotesize{
Field-theoretic description of Carrollian theories has largely remained classical so far. In this paper, we attempt to study the renormalization of Carrollian gauge field theories via path integral techniques. The case of Carrollian electrodynamics minimally coupled to a massive Carrollian scalar is considered. We report potential problems such as IR divergences and mass shell singularity cropping up at the first order in the perturbation. Perhaps, the most important result that we report is how conventional arguments for gauge independence for mass and coupling are invalidated for a gauge theory in a Carrollian setting. As of now, the renormalization of Carrollian gauge field theories seems to suffer from unphysical ramifications. Possible cures to resolve these issues are suggested. }
\end{abstract}	
\maketitle

\small

\section{Introduction}
\begin{figure}[h]
\centering
\includegraphics[scale=0.5]{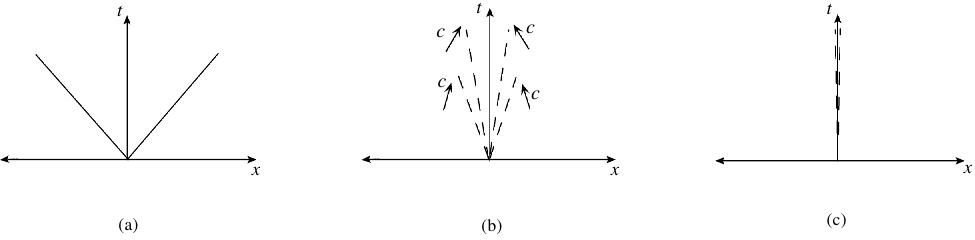}
\caption{\small{The figure in panel (a) is the light cone in Minkowski spacetime. Light travels along the path $x = ct$. In panel (b), we can see the light rays starts to collapse on the t axis as we approach closer and closer to the Carrollian limit. Finally the light cone collapse into $\displaystyle x=\lim_{c \to 0} ct \to 0$ in panel (c) above.}}
\label{fig:clight}
\end{figure}

Carrollian limit described as the speed of light $(c)$ going to zero was first introduced in \cite{Levy1965}\cite{1966ND} as a nontrivial contraction, as opposed to the well-known Galilean limit $(c \to \infty)$ of the Poincar\'{e} transformations. Owing to the deviation from the Lorentzian character, these two limits are also called non-Lorentzian limits. An illustrative way of understanding the Carrollian limit is the closing of the light cone to the time axis as depicted in \autoref{fig:clight}. A peculiar consequence of taking the Carrollian limit on the Poincar\'{e} transformation is that it renders the space as absolute i.e, not affected by boosts. Under such a setting causality almost disappears and the only way for two events to interact causally is if they happen at the same space and time point. For this very reason, the Carrollian limit is sometimes referred to as the ultra-local limit.  \\[5pt]
The last decade has seen a flurry of research activity in constructing field theories that are consistent with Carrollian symmetry (see \cite{Bagchi:2019clu}\cite{Banerjee:2020qjj}\cite{Baiguera:2022lsw}\cite{Chen:2023pqf}\cite{Bagchi:2022eav} and references therein). Carrollian symmetry is described by a set of symmetry generators viz. spatial and temporal translations, homogeneous rotations and Carrollian boosts. These symmetry generators can be obtained by taking $c \to 0$ limit of Poincar\'{e} symmetry generators. Equivalently, one may also wish to work in the natural system of units where $c$ is set to unity and rescale the space $(x_i)$ and time $(t)$ instead. The Carrollian limit is then defined as
\begin{equation*}
t \to \epsilon t \qquad, \qquad x_i \to x_i \qquad,\qquad \epsilon \to 0
\end{equation*}
which also leads to the Carrollian symmetry generators \cite{Bagchi:2019clu}\cite{Banerjee:2020qjj}.\\[5pt]
Over these years Carrollian symmetry has paved its way into many physics systems ranging from condensed matter\cite{Marsot:2022imf}\cite{Bagchi:2022eui} to black holes\cite{Donnay:2019jiz}. For example, it has been realized recently that a Carroll particle subjected to an external electromagnetic field mimics a Hall-type scenario\cite{Marsot:2022imf}. Furthermore, the emergence of Carrollian physics in the study of bi layer graphene\cite{Bagchi:2022eui}, the relation of Carrollian symmetry with plane gravitational waves\cite{Duval:2017els}, motion of particles on a black hole horizon\cite{Gray:2022svz} and hydrodynamics \cite{Bagchi:2023ysc}\cite{Ciambelli:2018wre} further fuels the need of Carrollian physics. In recent years, Carrollian holography has also emerged as a possible candidate for the flat space holography program \cite{Donnay:2022aba}\cite{Bagchi:2022emh}\cite{Bagchi:2016bcd}\cite{Duval:2014uva}. Some aspects of Carrollian gravity have also been studied in \cite{Henneaux:2021yzg}\cite{Henneaux:1979vn}\cite{Perez:2021abf}. \\[5pt]
However, much of the work carried out in the Carrollian sector has largely remained classical so far and not much heed has been paid to the quantization. As a matter of fact, the whole program of quantization of non-Lorentzian theories is fairly recent. For example, quantum studies on the Galilean field theories have surfaced in the last few years only (see \cite{Banerjee:2022uqj}\cite{Sharma:2023chs}\cite{Chapman:2020vtn}\cite{Baiguera:2022cbp}). This paper attempts to understand the quantum `nature', particularly,  the renormalization of Carrollian field theories.
\\
Understanding the quantum nature of Carrollian field theories is important on many levels. Firstly, as mentioned in the beginning, the Carrollian limit causes the light cone to close on the time axis and thus, time ordering is preserved only along the time axis. This results in two-point correlation functions of a Carrollian field theory to exhibit ultra-local behaviour at the tree level (see \autoref{section:prop}). 
It becomes intriguing to ask how Carrollian fields interact at the quantum level. Secondly, in the massless regime, certain Carrollian field theories at the classical level, admits invariance under the infinite conformal symmetries (for example \cite{Bagchi:2019clu}\cite{Banerjee:2020qjj}). It is then natural to ask whether these symmetries survive the quantization or not. Finally, it has been well established that the black hole horizon is a natural Carroll surface\cite{Donnay:2019jiz}. Thus, a quantum field theory living on the black hole horizon could be a Carrollian quantum field theory. \\[5pt]
In this paper, we have attempted to probe the renormalization of Carrollian electrodynamics\footnote{Here, by Carrollian electrodynamics we mean the electric sector of Carrollian electrodynamics\cite{Bagchi:2019clu}. In actuality, Carrollian electrodynamics also admits another sector known as the magnetic sector. For more details on the magnetic sector of Carrollian electrodynamics the reader is referred to \cite{Banerjee:2020qjj}.}\cite{Bagchi:2019clu} minimally coupled to a massive Carrollian scalar. At the classical level, the Lagrangian for the theory is obtained by Carroll limiting the massless Lorentzian scalar electrodynamics. The resulting theory consists of a gauge couplet $(B, A_i)$ minimally coupled to a complex scalar field $\phi$ through the coupling $e$. We then incorporate a mass term in the theory strictly constrained by the Carrollian symmetry. Owing to an interaction between gauge fields and a scalar field, we name the theory scalar Carrollian electrodynamics (sCED). To explore the renormalization description, we have made use of path integral techniques. We strictly restrict the renormalization scheme up to the first order in the perturbation i.e, 1 loop. Although the theory is renormalizable, there are serious unphysical ramifications, especially regarding the notion of mass and coupling in the Carrollian setting. The renormalization scheme leads to the notion of gauge-dependent mass and coupling which invalidates the conventional arguments of gauge independence for mass and coupling.  \\[5pt]
This paper is organized as follows: We have a total of 4 sections including the introduction. In \autoref{section:classical} we present the classical field description of sCED. A brief discussion on the Carrollian symmetry is presented followed by the Lagrangian formulation of sCED. Relevant Noether charges are constructed and it is shown that Carrollian algebra is satisfied at the level of charges. In \autoref{section:quantum} we proceed with the quantum field description of sCED. We propose path integral quantization and study renormalization of the theory up to 1 loop. Relevant results are then discussed and concluded in \autoref{section:conclusion}.

\section{Classical analysis of scalar Carrollian electrodynamics}
\label{section:classical}
\subsection{Carrollian Symmetry: A cursory visit!}
Carrollian symmetrty of a $(d+1)$ dimensional spacetime is described by time translations $(H)$, space translations $(P_i)$, homogeneous rotations $(J_{ij})$ and Carrollian boosts $(B_i)$. In an adaptive coordinate chart $x^I =(t,x^i)$ we can express them as
\begin{equation}
\label{eqn:generators}
H =\partial_t\;\;,\;P_i =\partial_i \;\;,\;\; J_{ij}=(x_i \partial_j-x_j \partial_i) \;\;,\;\;B_i=x_i \partial_t,
\end{equation}
The symmetry generators (\ref{eqn:generators}) can be obtained by Carroll limiting the Poincar\'{e} symmetry generators \cite{Bagchi:2019clu}\cite{Banerjee:2020qjj}\cite{Bagchi:2019xfx}. However, there also exists a yet another way i.e, geometric way to arrive at the Carrollian symmetry generators (see for example \cite{Duval:2014uoa} or \autoref{section:geometry}). The symmetry generators (\ref{eqn:generators}) form a closed Lie algebra called Carrollian algebra given by
\begin{eqnarray}
&[J_{ij}, B_k ]&=\delta_{k[j}B_{i]} \;\;, \;\; [J_{ij}, P_k ]=\delta_{k[j}P_{i]} \;\;,\;\;[J_{ij}, H ]=0  \nonumber \\
&[B_i,P_j]&=-\delta_{ij}H \;\;,\;\;[P_i,H]=0\nonumber\\
\label{eqn:algebra}
&[P_i,P_j]&=0 \;\;,\;\;[B_i,H]=0
\end{eqnarray}
The generators $\{H,P_i,J_{ij},B_i\}$ can be used to study the action of symmetry generators on the fields at a general spacetime point i.e, for a generic scalar field $\varphi$ and a generic vector field $V_i$, we can write (see \cite{Bagchi:2019clu} and references therein for complete details) 
\begin{equation}
\label{eqn:fieldaction}
\begin{split}
&\text{Spatial rotations: }\delta_{\omega} \varphi(t,x) =\omega^{ij} (x_{[i} \p_{j]}) \varphi(t,x) \\
&\qquad \qquad \qquad \qquad \delta_{\omega} V_l(t,x)=\omega^{ij} \big[(x_{[i} \p_{j]}) V_l(t,x)+\delta_{l[i}V_{j]}\big]\\[5pt]
&\text{Carrollian boosts: }\delta_{\sss B} \varphi(t,x) =b^j[x_j\p_t \varphi(t,x)]\\
&\qquad \quad \qquad \qquad \quad \delta_{B}V_l(t,x)=b^j \big[x_j\p_t V_l(t,x)+ \delta_{lj} \varphi(t,x) \big]\\[5pt]
&\text{Space translation: }\delta_{p} \varphi(t,x) =p^j \p_j \varphi(t,x)\\
&\qquad \qquad \qquad \qquad \delta_{p} V_i(t,x) =p^j \p_j V_i(t,x)\\[5pt]
&\text{Time translation: }\delta_{H} \varphi(t,x) =\partial_t \varphi(t,x)\\
&\qquad \qquad \qquad \qquad \delta_{H} V_i(t,x) =\partial_t V_i(t,x)\\[5pt]
\end{split}
\end{equation}
where $\omega^{ij}$ is an antisymmetric matrix and $b^i$ and $p^i$ are the boosts and spatial translation parameters. We shall be employing (\ref{eqn:fieldaction}) to demonstrate the invariance of sCED and then later again to construct the conserved charges associated to these symmetry generators for sCED.

\subsection{Lagrangian and conserved charges for sCED}
We begin our discussion by proposing the Lagrangian for massive sCED. It must be noted that the Lagrangian for the massless scalar Carrollian electrdoyanmics was propsed in  \cite{Bagchi:2019clu}. Their technique relied on Helmholtz integrability conditions\footnote{Helmholtz conditions are the necessary and sufficient conditions which when satisfied by a set of second order partial differential equations, guarantees an \emph{action}. We request the reader to check \cite{Bagchi:2019clu}\cite{Banerjee:2020qjj}\cite{10.2307/1989912} for more details on the method and its applications.}  In a coordinate chart $x^I=(t,x^i)$ the Carroll invariant Lagrangian $\mathcal{\tilde{L}}$ for massless sCED is given by,
\begin{equation}
\mathcal{\tilde{L}}=\frac{1}{2} \Big \{(\partial_i B)^2+(\partial_t A_i)^2 -2 (\partial_t B) (\partial_i A_i)\Big\}-(D_t \phi)^* (D_t \phi)
\end{equation}
where
$D_t \phi= \partial_t\phi+ie B \phi$ and $(D_t \phi)^*=\partial_t {\phi}^*-ie B \phi^*$. We add a mass term strictly constrained by the Carrollian symmetry (\ref{eqn:fieldaction}) to the above Lagrangian such that the Lagrangian $\mathcal{L}$ for massive sCED\footnote{From here onwards, we shall simply call massive scalar Carrollian electrodyanmics as sCED.}\;  is given by,
\begin{equation}
\label{eqn:lag}
\begin{split}
\mathcal{L}&=\frac{1}{2} \Big \{(\partial_i B)^2+(\partial_t A_i)^2 -2 (\partial_t B) (\partial_i A_i)\Big\}-(\partial_t \phi^*)(\partial_t \phi) \\
&+ m^2 \phi^* \phi -i e B \Big[\phi \partial_t \phi^*-\phi^* \partial_t \phi\Big]-e^2 B^2 \phi^* \phi
\end{split}
\end{equation}
The equations of motion for sCED can be obtained by varying (\ref{eqn:lag}) with respect to the fields $B,A_i$ and $\phi$, resulting in:
\begin{eqnarray}
&\partial_t\partial_t A_i-\partial_t\partial_i B&=0 \nonumber\\
\label{eqn:eom}
&D_t D_t \phi+m^2\phi&=0\\
&\partial_i\partial_t A_i-\partial_i\partial_i B&-ie\(\phi D_t^* \phi^* - \phi^* D_t \phi\)=0 \nonumber.
\end{eqnarray}
which agrees with \cite{Bagchi:2019clu} if we set $m=0$ in (\ref{eqn:eom}). It is instructive to note that the Lagrangian (\ref{eqn:lag}) enjoys the following gauge invariance:
\begin{eqnarray}
&\delta_\alpha B&= \alpha_1\\
&\delta_\alpha A_i&= -\partial_i \alpha_2
\end{eqnarray}
where $\alpha_1$ and $\alpha_2$ are arbitrary functions (For a detailed discussion on the gauge structure of Carrollian electrodynamics we direct the reader to \autoref{section:dirac}).\\[5pt]
Noether theorem suggests that associated to every continuous symmetry of the Lagrangian, there exists a corresponding global conserved charge. Since the Lagrangian (\ref{eqn:lag}) is invariant under Carrollian symmetry \eqref{eqn:fieldaction}, the associated Noether charges for rotations $(Q(\omega))$, space and time translations $(Q(p), Q(h))$ and boosts $(Q(b))$ are given by

\bes{}
\label{char1}
\bea{}
\begin{split}
&Q(\omega)=\int d^{d-1}x ~\omega^{ij}   \lb \dot{A}_k\(x_{[i}\p_{j]}A_k+\delta_{k[i}A_{j]}\)\\
&-(\p \cdot A)\(x_{[i}\p_{j]}B\) -(D_t \phi)^*(x_{[i}\p_{j]}\phi)- (x_{[i}\p_{j]}\phi)^* (D_t \phi)  \rb \nonumber\\
&Q(p)=\int d^{d-1}x ~ p^l\lb \dot{A}_i \p_l A_i- \p_l B \, \p \cdot A\\
&\qquad \qquad \qquad \qquad \qquad \quad -(D_t \phi)^* \p_l \phi-\p_l \phi^* D_t\phi \rb \nonumber\\
&~Q(h)=\int d^{d-1}x ~ \lb  \frac{1}{2}(\dot{A}_i^2-(\p_i B)^2)+ (D_t \phi)^* (D_t \phi)\\
&\qquad \qquad \quad-(D_t \phi)^* (\p_t \phi)-(\p_t \phi)^* (D_t \phi)-m^2\phi\phi^{*} \rb \nonumber\\
&~Q(b)=\int d^{d-1}x ~b^l x_l  \lb  \frac{1}{2}(\dot{A}_i^2-(\p_i B)^2)+ (D_t \phi)^* (D_t \phi)\\
&\qquad -(D_t \phi)^* (\p_t \phi) -(\p_t \phi)^* (D_t \phi)-m^2\phi\phi^{*} \rb+b^l (\dot{A}_l B)\nonumber
\end{split}
\eea\ees
Correspondingly, after a bit of lengthy but straightforward calculation we can arrive at the charge algebra. The non vanishing Poisson brackets for sCED are,
\begin{eqnarray*}
\{Q(\omega), Q(p)\} = Q(\tilde{p})\\
\{Q(\omega), Q(b)\} =Q(\tilde{b}) \\
\{Q(p), Q(b)\} = Q(h)
\end{eqnarray*}
where $\tilde{p} \equiv \tilde{p}^{k}\partial_k=\omega^{ij}p_{[j}\partial_{i]}$ and $\tilde{b} \equiv \tilde{b}^{k}\partial_k=\omega^{ij}b_{[j}\partial_{i]}$. Clearly the Carrollian algebra is realized at the level of Noether charge algebra. We are now in position to probe into the quantum field description for sCED.

\section{Quantum field description of sCED}
\label{section:quantum}
In the previous section we studied the classical field description of scalar Carrollian electrodynamics. In this section, we propose a quantization prescription, particularly the renormalization of sCED\footnote{It should be noted that the quantization of Carrollian field theory is not on a firm footing and we are working on addressing the canonical quantization of Carrollian field theories. We shall be reporting these issues with glorifying detail in our upcoming work (the manuscript is currently under preparation). However, for completeness, we make a very generic and plausible assumption of the existence of the vacuum and present a cursory introduction to the renormalization of an interacting (quartic) Carrollian scalar field theory in \autoref{section:CSP} which makes the renormalization approach for sCED self-sufficient.} We shall put to use functional techniques to explore the renormalization of scalar Carrollian electrodynamics. The \emph{action} $S$, for the sCED using \eqref{eqn:lag} takes the following form
\begin{equation}
\label{eqn:action}
\begin{split}
S&=\bigintsss dt d^3x\; \Bigg[\frac{1}{2} \Big \{(\partial_i B)^2+(\partial_t A_i)^2 -2 (\partial_t B) (\partial_i A_i)\Big\}\\
&-(\partial_t \phi^*)(\partial_t \phi)-i e B \Big[\phi \partial_t \phi^*-\phi^* \partial_t \phi\Big]-e^2 B^2 \phi^* \phi + m^2 \phi^* \phi \Bigg]
\end{split}
\end{equation}
The gauge field couplet $\varphi^I \equiv (B,A^i)$ and the complex scalar field $\phi$, carries the mass dimensions $[B]=[A_i] =[\phi]=[\phi^{*}]=1$ rendering us with a case of a marginally renormalizable theory with $[e]=0$. An instructive thing to note in (\ref{eqn:action}) is that the gauge field $A_i$ does not participate in any interaction with $\phi$ or $\phi^*$. As a consequence, the propagators and vertices shall admit loop corrections offered only due to the interaction between the gauge field $B$ and the complex scalar $\phi$. For the rest of the paper, we shall focus only on the 1 loop corrections in the theory.

\subsection{Feynman Rules}
Since sCED is a gauge theory \footnote{The gauge structure of sCED is because of the gauge couplet $(B, A_i)$ in the theory. To understand its gauge structure in more detail please refer to \autoref{section:dirac}.} it is important that we gauge fix the theory. We shall employ the gauge fixing technique developed by Faddeev and Popov \cite{ryder_1996}\cite{Peskin:1995ev}\cite{FADDEEV196729} i.e, the gauge fixed action is
\begin{equation}
\label{eqn:gact1}
\begin{split}
S&=\bigintsss dt d^3x\; \Bigg[\frac{1}{2} \Big \{(\partial_i B)^2+(\partial_t A_i)^2 -2 (\partial_t B) (\partial_i A_i)\Big\}\\
&\quad -(\partial_t \phi^*)(\partial_t \phi) -i e B \Big[\phi \partial_t \phi^*-\phi^* \partial_t \phi\Big]-e^2 B^2 \phi^* \phi\\
&\qquad \qquad \qquad + m^2 \phi^* \phi \Bigg] +\bigintsss dt d^3 x\; \mathcal{L}_{\text{gauge fixed}}
\end{split}
\end{equation}
with $\mathcal{L}_{\text{gauge fixed}}$ given by
\begin{equation*}
\mathcal{L}_{\text{gauge fixed}} = -\frac{1}{2 \xi} \bigg(G[B(t,x^i), A^i(t,x^i)]\bigg)^2
\end{equation*}
\linebreak
where $G[B,A^i]$ is the gauge fixing condition and $\xi$ is the gauge fixing parameter. \\[5pt]
We choose $G[B,A^i] = (\partial_t B)$ such that the gauge fixed action (\ref{eqn:gact1}) becomes
\begin{equation}
\label{eqn:gact2}
\begin{split}
S&=\bigintsss dt d^3x\; \Bigg[\frac{1}{2} \Big \{(\partial_i B)^2+(\partial_t A_i)^2 -2 (\partial_t B) (\partial_i A_i)\Big\} \\
&\qquad-(\partial_t \phi^*)(\partial_t \phi) -i e B \Big[\phi \partial_t \phi^*-\phi^* \partial_t \phi\Big]-e^2 B^2 \phi^* \phi \\
&\qquad \qquad+ m^2 \phi^* \phi  -\frac{1}{2 \xi} (\partial_t B)^2 \Bigg]
\end{split}
\end{equation}
Observe that we can arrive at the same gauge fixed action by Carroll limiting the Lorentz gauge fixing condition for Lorentzian scalar Electrodynamics. Also, notice that we have omitted the Fadeev-Popov ghost term in (\ref{eqn:gact2}). This is because the Faddeev-Popov ghosts does not interact with the gauge field couplet $(B,A_i)$ and hence does not contribute to any of the loop corrections.\\[5pt]
Now, with the gauge fixed action (\ref{eqn:gact2}) at our disposal, we can evaluate the propagator for the gauge couplet $\varphi^I$. For the sake of brevity, we introduce $\bm{p}$=$(\omega, p_i)$ such that the gauge field propagator $D_{IJ} =\big< \varphi_I, \varphi_J \big>$ reads, 
\begin{equation}
\label{eqn:matprop}
D_{IJ}=-i\begin{pmatrix}
\dfrac{\xi}{\omega^2}\qquad & \dfrac{\xi}{\omega^3} p_i\\[20pt]
\;\dfrac{\xi}{\omega^3} p_i\qquad & \;\;\;-\dfrac{\delta_{ij}}{\omega^2}+ \dfrac{p_i p_j}{\omega^4} \xi\;
\end{pmatrix}
\end{equation}\\
and the propagator for the complex scalar field $\phi$ takes the following form
\begin{equation}
\label{eqn:sprop}
\big<\phi,\phi^*\big> = \frac{i}{-\omega^2+m^2}
\end{equation}
Before we proceed further, notice that the gauge field propagator (\ref{eqn:matprop}) admits a pole at $\omega =0$ which essentially captures the ultra local behaviour of Carrollian field theories i.e, two events are causally related to each other only if they happen at the same spacetime point. This can be confirmed further by Fourier transforming the propagator in position space (see \autoref{section:prop} for more details). A similar feature can be observed for the complex scalar field propagator (\ref{eqn:sprop}). However, it must be noted that \eqref{eqn:sprop} admits a pole at $\omega^2=m^2$, which is precisely how mass is defined for a free theory under quantum field theory setting \cite{ryder_1996}\cite{Peskin:1995ev}.\\[5pt] 
The Feynman rules for sCED are then given by
\begin{equation}
\label{eqn:feynrule}
\begin{split}
&1. \qquad \text{Gauge scalar propagator,} \qquad \Big\<B\;\;,\;\;B \Big\>\;\;=\;\;\dfrac{-i}{\omega^2} \xi\\[4pt]
& 2.  \qquad \text{Scalar propagator,} \qquad \qquad \quad \Big\<\phi^*\;\;,\;\;\phi\Big\>\;\;=\;\; \dfrac{i}{-\omega^2+m^2}\\[4pt]
&3.  \qquad \text{Three point vertex,} \qquad \quad \;\; \quad V_{\tiny{B \phi^* \phi}}= \;\;i e (\omega_p-\omega_q)\\[4pt]
&4.  \qquad \text{Four point vertex,} \qquad \qquad \quad V_{\tiny{B^2 \phi^* \phi}}= \;\; -2 i e^2\\[4pt]
\end{split}
\end{equation}
The diagrammatic  representation of (\ref{eqn:feynrule}) is given in \autoref{fig:feynt}.
\begin{table}
\begin{center}
\begin{tabular} { | m{1cm} | m{3cm} | m{4cm} | }
\hline
1. & $\qquad \;\; \Big<B, B\Big>$ &\qquad \includegraphics[scale=0.3]{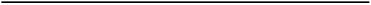}\qquad \;\\[7pt]
\hline
2. & \qquad \; $\Big<\phi, \phi^*\Big>$ &\qquad \includegraphics[scale=0.35]{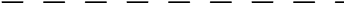}\qquad \;\\[7pt]
\hline
3. & \qquad \quad $V_{\tiny{B \phi^* \phi}}$ &\qquad \includegraphics[scale=0.35]{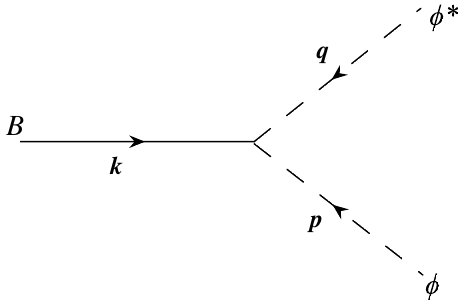}\qquad\\[7pt]
\hline
4. & \qquad \quad $V_{\tiny{B^2 \phi^* \phi}}$ & \quad \includegraphics[scale=0.35]{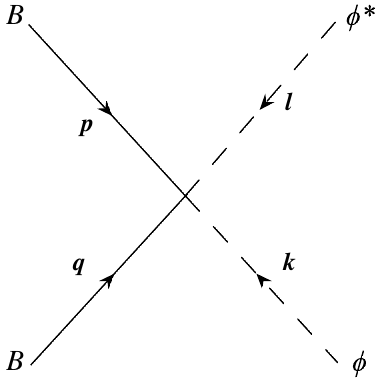}\qquad\\
\hline
\end{tabular}
\caption{\label{fig:feynt}Feynman rules for sCED}
\end{center}
\end{table}
Notice that we have purposefully omitted the propagators $\big< B,A_i\big>$ and $\big<A_i, A_j\big>$ while writing down the \eqref{eqn:feynrule}. This is because the only allowed interaction in the theory is between the fields $B$ and $\phi$ (and its complex conjugate) and thus $\big< B,A_i\big>$ and $\big<A_i, A_j\big>$ will not contribute to any loop corrections in the theory. In what follows, we shall evaluate the necessary 1 loop corrections to the propagators and vertices.

\subsection{Renormalization}
Owing to the 3-point and 4-point interactions between the gauge field $B$ and the complex scalar $\phi^*$ and $\phi$, the theory of sCED admits 1 loop corrections to the propagators and the vertices. Generally, these loop integrals diverge at large values of energy ($\omega$) and momentum ($|p|$) and lead to what is known as UV divergences. In order to make sense of these divergent integrals, we employ the technique of cut-off regularization where we set an upper cutoff, $\Omega$ in the energy sector and $\Lambda$ in the momentum sector. In addition to UV divergences, the loop integrals may also diverge at low energy (or momentum) scales. This is called IR divergence. Most often, such divergences are encountered in massless theories where the pole of the propagator admits a mass-shell singularity. It is important to realize that the gauge propagator for sCED (\ref{eqn:feynrule}) showcases a similar pole structure. Thus some of the loop corrections shall admit IR divergences. However, physical observables such as correlation functions shall not depend on IR divergences. This essentially means that renormalized gauge propagator should not contain any IR divergence. Interestingly, we shall see later that under the renormalization scheme, the gauge field propagator $\big<B,B\big>$ does not admit any IR divergence\footnote{It must be pointed out that the propagator for the complex scalar field is not gauge invariant and hence, its renormalization may depend on the gauge fixing parameter $\xi$.}\\[5pt]
For the present discussion, we are concerned with the renormalization of sCED, hence we shall only retain UV divergent terms and ignore IR divergences. But before we proceed any further, we shall comment on the issue of impromptu ignoring IR divergences.  Recall that in Lorentzian quantum electrodynamics (QED), IR divergences are handled by the inclusion of soft photons of mass $\mu$ such that, in the limit $\mu \to 0$ IR divergences neatly cancels. This technique does not hold for the case of sCED.
A similar problem of IR divergences occurs in the study of scattering amplitude for non-relativistic QED, where ignoring the IR divergences at the first few orders of the perturbation leads to the correct results\cite{Caswell:1985ui}\cite{Labelle:1996en}. Lastly, the problem of IR divergences has also been observed for the case of scalar Galilean electrodynamics (sGED)\cite{Chapman:2020vtn} where ignoring IR divergences leads to a renormalized theory of sGED. For the rest of the discussion, we shall abide by this approach and plan to examine the resolution of IR divergences in the Carrollian setting in the future. 

\subsubsection{Loop corrections and renormalization conditions}
\begin{figure}[h]
\centering
\includegraphics[scale=0.5]{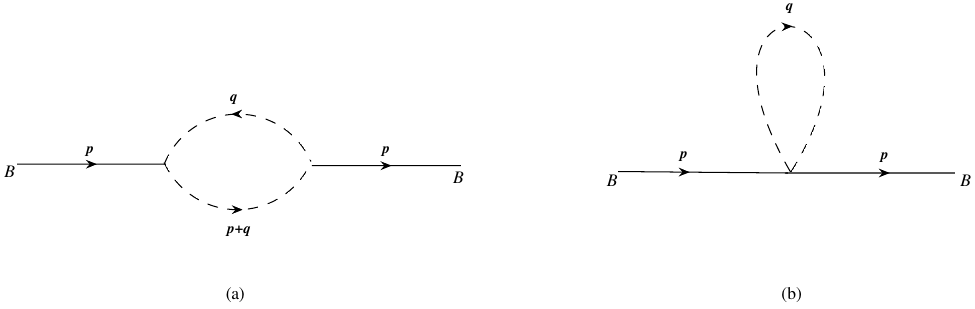}
\caption{In this panel, diagram (a) is the three point correction to the $\big<B,B\big>$ propagator and diagram (b) is the four point correction to the $\big<B,B\big>$ propagator.}
\label{fig:corrB}
\end{figure}

\begin{figure}[h]
\centering
\includegraphics[scale=0.5]{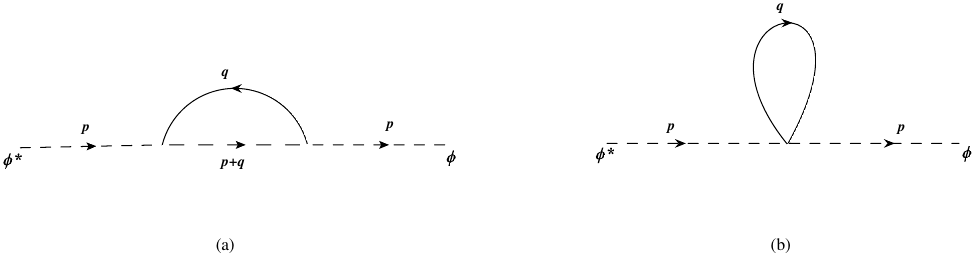}
\caption{In this panel, diagram (a) is the three point correction to the $\big<\phi^*,\phi\big>$ propagator and diagram (b) is the four point correction to the $\big<\phi^*,\phi\big>$ propagator.}
\label{fig:corrS}
\end{figure}

\begin{figure}[h]
\centering
\includegraphics[scale=0.4]{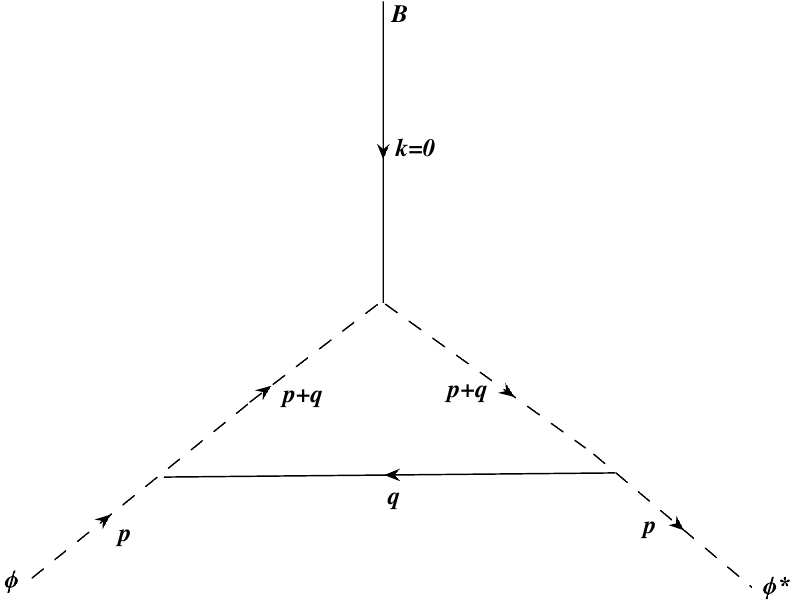}
\caption{Correction to the three point vertex $V_{B \phi^* \phi}$}
\label{fig:corrV3}
\end{figure}

The two propagators we are interested in are gauge field propagator $\big< B,B\>$ and the complex scalar propagator $\big< \phi, \phi^*\big>$. We shall begin our discussion with the gauge field propagator $\big< B, B\big>$. The relevant 1 loop corrections are drawn in \autoref{fig:corrB}.
\\
The loop correction $(\Sigma_1)$ offered to the $\big< B, B\big>$ propagator due to $V_{B\phi^* \phi}$  can be evaluated by integrating along unconstrained variable $(\omega_q, q)$ of diagram (a) in \autoref{fig:corrB} i.e, 
\begin{equation}
\Sigma_1= \bigintsss  d\omega_q  d^3 q\;  \frac{e^2 (2 \omega_q+\omega_p)^2}{(-\omega_q^2+m^2)(m^2-(\omega_q+\omega_p)^2)}
\end{equation}
The superficial degree of divergence suggests that the integral converges in the energy sector but diverges cubically at large values of $q$. To this end, we put a UV cut-off $\Lambda$ in the momentum sector. Also, it must be observed that the integral does not contain any IR divergence since the integrand is well defined at $\omega_q \to 0$. A straight forward calculation then gives
\begin{equation}
\label{eqn:gcor3}
\Sigma_1 =i\frac{8\pi^2 e^2 \Lambda^3}{3 m}
\end{equation}
Notice that the degree of divergence of $\Sigma_1$ is cubic which agrees with the predicted degree of divergence.  Next, we shall evaluate the correction $(\Sigma_2)$ offered due to $V_{B^2 \phi^* \phi}$. The Feynman diagram is given in diagram (b) of \autoref{fig:corrB}. The integral $\Sigma_2$ reads
\begin{equation}
\Sigma_2= \bigintsss d\omega_q d^3q\; \frac{2 e^2}{(m^2-\omega_q^2)}
\end{equation}
As before, the integral diverges cubically at large values of $q$ but remains convergent in $\omega_q$. The integral evaluates to
\begin{equation}
\label{eqn:gcor4}
\Sigma_2= -i\frac{8\pi^2 e^2 \Lambda^3}{3 m}
\end{equation}
With (\ref{eqn:gcor3}) and (\ref{eqn:gcor4}) at are disposal, the propagator $\big<B, B\big>$ upto first order in the perturbation i.e, $\mathcal{O}(e^2)$ is given by
\begin{equation*}
\includegraphics[scale=0.5]{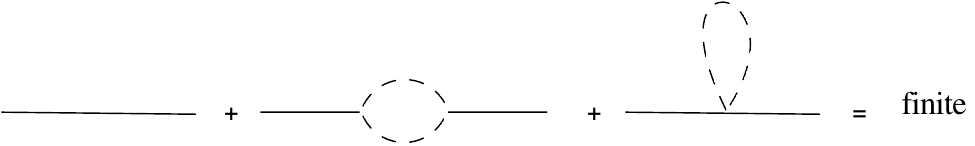}
\end{equation*}
Mathematically, we can write
\begin{equation*}
\begin{split}
&-\frac{i \xi}{\omega^2}+ \bigg(-\frac{i \xi}{\omega^2}\bigg) \Bigg[i\frac{8\pi^2 e^2 \Lambda^3}{3 m}\Bigg]\bigg(-\frac{i \xi}{\omega^2}\bigg)\\
&+\bigg(-\frac{i \xi}{\omega^2}\bigg)\Bigg[-i\frac{8\pi^2 e^2 \Lambda^3}{3 m}\Bigg] \bigg(-\frac{i \xi}{\omega^2}\bigg) = \text{finite}
\end{split}
\end{equation*}
Since the contribution from the three-point correction exactly cancels the contribution from the four-point correction, we end up with a finite value, which essentially means that to the order $\mathcal{O}(e^2)$ in the perturbation the gauge field propagator $\Big< B, B,\Big>$ remains finite and does not require any counter term. This allows us to make a redefinition $B_{(b)} =B$, where the subscript $b$, represents the bare field. Also recall that there is no interaction allowed for the vector field $A^i$ in the theory \eqref{eqn:gact2} which essentially means that the gauge field $B$ and $A_i$ follow the field redefinitions:
\begin{eqnarray}
\label{eqn:gaugerenor}
&B_{(b)}& = B\\
& A^i_{(b)}&= A^i
\end{eqnarray}
We now turn our attention to 1 loop corrections to the $\big< \phi^*, \phi\big>$. The allowed Feynman diagrams are given in \autoref{fig:corrS}.
\\
The expression for the loop integral ($\Pi_1$) in diagram (a) of \autoref{fig:corrS} takes the following form
\begin{equation}
\Pi_1= \bigintsss d\omega_q d^3q\; \frac{2 e^2 \xi (\omega_q+2 \omega_p)^2}{\omega_q^2 (m^2-(\omega_q+ \omega_p)^2)}
\end{equation}
As before, the integral diverges cubically at large value of $q$ and thus we put a UV cut off $\Lambda$ in the momentum sector. In addition, the integral also admits an IR divergence. The source of the IR divergence is the mass shell singularity present in the pole structure of the gauge field propagator and thus the integrand diverges at $\omega_q \to 0$. As already discussed, we shall ignore the IR divergence piece and retain only the UV divergent part of the integral. The integral evaluates to 
\begin{equation}
\label{eqn:phicorr3}
\Pi_1= -i\frac{8 \pi^2 e^2 \xi \Lambda^3}{3} \Bigg[\frac{1}{m}+\frac{8 m \omega_p^2}{(m^2-\omega_p^2)^2} \Bigg]
\end{equation}
Finally, the loop integral $(\Pi_2)$ in diagram (b) of \autoref{fig:corrS} reads
\begin{equation}
\Pi_2 =-2 e^2 \xi \bigintsss d\omega_q d^3 q\; \frac{1}{\omega_q^2}
\end{equation}
Clearly, the integrand diverges at $\omega_q \to 0$ leading to an IR divergent piece which along the previous lines shall be ignored. Thus, the only UV divergent piece we have is $\Pi_1$ which we shall be able to absorb by introducing the counter term i.e, 
\begin{equation*}
\includegraphics[scale=0.53]{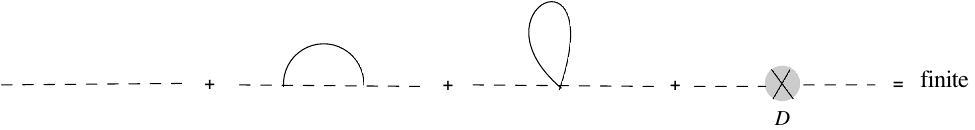}
\end{equation*}
where the last term is the counter term that we have added with $D$ as its coefficient. Mathematically, we can then write down\footnote{Note that for notational agreement, $\omega_p$ is now denoted by $\omega$.}
\begin{equation}
\label{eqn:renorcondmass}
\frac{i}{-\omega^2+m^2-i(D-i e^2  f(\xi, m,\omega, \Lambda))} =\text{finite}
\end{equation}
where 
\begin{equation}
\label{eqn:call}
f(\xi, m,\omega, \Lambda) =\frac{8 \pi^2 \xi \Lambda^3}{3} \Bigg[\frac{1}{m}+\frac{8 m \omega^2}{(m^2-\omega^2)^2} \Bigg] 
\end{equation}
such that (\ref{eqn:renorcondmass}) leads to a finite value for
\begin{equation*}
D= i e^2  f(\xi, m,\omega, \Lambda) = ie^2 \frac{8 \pi^2 \xi \Lambda^3}{3} \Bigg[\frac{1}{m}+\frac{8 m \omega^2}{(m^2-\omega^2)^2} \Bigg] 
\end{equation*}
Notice that the mass dimensions of $D$ is 2 i.e, $[D]=2$, which essentially means that the pole of (\ref{eqn:renorcondmass}) defines the mass renormalization condition for sCED. We can then write
\begin{equation*}
D= i \delta m^2
\end{equation*}
where,
\begin{equation}
\label{eqn:renormass}
\delta m^2 =e^2 \frac{8 \pi^2 \xi \Lambda^3}{3} \Bigg[\frac{1}{m}+\frac{8 m \omega^2}{(m^2-\omega^2)^2} \Bigg] 
\end{equation}
Clearly, the corresponding counter term in the Lagrangian is
\begin{equation}
\label{eqn:mLagcounter}
(\mathcal{L}_{ct})_1 = \delta m^2 \phi^* \phi
\end{equation}
Although we managed to absorb the divergences via counter term (\ref{eqn:mLagcounter}), there is something very unsettling about it. Notice that $\delta m^2$ depends upon gauge parameter $\xi$. This is unphysical, for mass should remain independent of the choice of gauge parameter. In fact, it is not just about the mass, even coupling turns out to depend on $\xi$ upon renormalization. This can be demonstrated by carrying out the renormalization for three point vertex $V_{B \phi^* \phi}$. The only possible correction to the vertex is given in \autoref{fig:corrV3}. 
\\
Following the renormalization scheme we can check that the counter term needed to absorb the divergences for the three point vertex is
\begin{equation}
\label{eqn:eLagcounter}
(\mathcal{L}_{ct})_2= -G \;i e B (\phi \partial_t \phi^*-\phi^* \partial_t \phi)
\end{equation}
where
\begin{equation}
\label{eqn:renormcoup}
G=-\frac{8 \pi^2 e^2 \xi \Lambda^3}{3 (m^2-\omega^2)} \Bigg[\frac{1}{m}+\frac{8 m \omega^2}{(m^2-\omega^2)^2}\Bigg]
\end{equation}
is the renormalization coefficient and evidently depends on the the gauge fixing parameter $\xi$. The procedure of absorbing the UV divergent terms is not unique in quantum field theory. It is instructive to note here that in the absence of counter terms, the role of the correction (\ref{eqn:phicorr3}) is to shift the mass $m$ (appearing in the Lagrangian) to the physical (renormalized) mass $m_{\tiny{\text{phy}}}$. Physically, this is interpreted as- mass $m$ is infinite and it takes infinite shift to bring it down to  $m_{\tiny{\text{phy}}}$ i.e, devoid of the counter term, the mass renormalization condition using (\ref{eqn:renorcondmass}) is given by,
\begin{equation*}
-\omega^2+m^2-e^2  f(\xi, m,\omega, \Lambda))\Bigg{|}_{\omega^2 =m^2_{\tiny{\text{phy}}}} =0
\end{equation*}
which implies 
\begin{equation}
\label{eqn:phymass}
m^2_{\tiny{\text{phy}}}= m^2-e^2 f
\end{equation}
where $f$ is given by (\ref{eqn:call}). However, an interesting thing to note here is that $m_{\tiny{\text{phy}}}$ is heavily gauged. For any physical theory, $m^2_{\tiny{\text{phy}}}$ should remain independent of the gauge fixing parameter. A similar calculation when carried out for the coupling leads to the same arguments. This invalidates the conventional arguments of gauge independence of mass and coupling. For any physical theory, the physical observables such as mass or coupling should not depend upon the choice of gauge parameter $\xi$. For example, in Lorentzian QED, it does not matter whether we work in the Feynman gauge $(\xi =1)$ or Landau gauge $(\xi =0)$, the coupling of the theory which is t\^{e}te-\`a-t\^{e}te related to the fine structure constant remains independent of the gauge choice.\\[5pt]
The occurrence of $\xi$ in the renormalization coefficients $(\delta m^2, G)$ renders an ambiguity in the definitions of mass and coupling strength. However, this ambiguity is not new in the quantum field theory arena. As a matter of fact, such behaviour has been observed in Lorentz's invariant quantum field theories as well. For example, in the massive Schwinger model in $(1+1)$ dimensions, the presence of mass shell singularities is known to invalidate the standard requirement for gauge independence of renormalized mass \cite{Das:2012qz}\cite{Das:2013vua}.  In the case of the Schwinger model, these ambiguities are resolved by using Nielsen identities which requires one to formulate the Lagrangian in `physical' gauge\footnote{Physical gauge refers to a gauge choice where the unphysical degree of freedom such as Faddeev-Popov ghosts decouple from a theory. For example axial gauge and Coulomb gauge. An advantage of working in physical gauges is that IR divergences are often softer and neatly separated.}\; \cite{Das:2012qz} \cite{Das:2013iha}. Nielsen identities provide a useful way to construct a notion of gauge independent renormalized mass \cite{Nielsen:1975fs} \cite{Breckenridge:1994gs}. It should be noted that the pole structure of the propagator for sCED shares a massive similarity to the Schwinger model in $1+1$ dimensions. However, to resolve the issue of gauge dependence of mass in sCED, we first need to formulate the Lagrangian in a physical gauge. One shortcoming of working with physical gauges such as axial gauges is that it does not fix the gauge completely and thus leaves a residual gauge degree of freedom. However, as far as gauge invariant quantities are concerned, it shall not matter what gauge we work with. Obviously, mass and coupling strength are the physical observables in a theory and should thus remain independent of the gauge choice. With the aim of resolving these ambiguities, it shall be interesting to study the quantization of sCED in this framework. We plan to address this problem in detail in the future.\\

\subsection{Counter term and bare Lagrangian}
In the preceding section we realized that the renormalized mass and coupling admits ambiguities, for they turn out to depend upon gauge parameter $\xi$. However, as already mentioned, the renormalization scheme is not unique. One of the way to counter off the UV divergences in the theory is to adhere to the method of counter terms. We conclude from the renormalization of three point vertex and propagators that the complex scalar field enjoys the following field redefinitions
\begin{equation}
\label{eqn:comrenor}
\phi_{(b)} =\phi \qquad \implies \quad \phi^*_{(b)} =\phi^*
\end{equation}
It then follows from (\ref{eqn:mLagcounter}) and (\ref{eqn:comrenor}) that bare mass term in the Lagrangian i.e, $\mathcal{L}_{(b)}^{(mass)}$ is given by
\begin{eqnarray}
&\mathcal{L}_{(b)}^{(mass)}& = m^2 \phi^* \phi + \delta m^2 \phi^* \phi \nonumber \\[5pt] 
\implies &\mathcal{L}_{(b)}^{(mass)}& =\big(m^2 +\delta m^2 \big) \phi^*_{(b)} \phi_{(b)} \nonumber \\[5pt]
\label{eqn:baremass}
\implies &\mathcal{L}_{(b)}^{(mass)}& = m^2_{(b)}  \phi^*_{(b)} \phi_{(b)}
\end{eqnarray}
where $m^2_{(b)}= \big(m^2 +\delta m^2\big)$ defines the bare mass of the theory. Similarly using \eqref{eqn:gaugerenor}, (\ref{eqn:eLagcounter}) and \eqref{eqn:comrenor} we can write the bare coupling term $\mathcal{L}_{(b)}^{(coupling)}$ as
\begin{eqnarray}
\label{eqn:barecoup}
\mathcal{L}_{(b)}^{(coupling)}& =- i e_{(b)} B_{(b)} \big (\phi_{(b)} \partial_t \phi^*_{(b)}-\phi^*_{(b)} \partial_t \phi_{(b)} \big)
\end{eqnarray}
where $e_{(b)}=e (1+G)$ defines the bare coupling in the theory. Lastly, demanding the consistency of field and coupling redefinition we can write down bare term involving four point interaction i.e, $\mathcal{L}_{(b)}^{quartic}$
\begin{eqnarray}
&\mathcal{L}_{(b)}^{quartic}& = -e^2 B^2 \phi^* \phi-\alpha^2e^2 B^2 \phi^* \phi \nonumber\\[5pt]
\label{eqn:barecoup1}
\implies &\mathcal{L}_{(b)}^{quartic}& = -e^2_{(b)} B^2_{(b)} \phi^*_{(b)} \phi_{(b)}
\end{eqnarray}
where $\alpha^2 = G(2+G)$. Finally, the bare Lagrangian $\mathcal{L}_{(b)}$ follows from (\ref{eqn:gaugerenor}), \eqref{eqn:comrenor}, \eqref{eqn:baremass}, \eqref{eqn:barecoup} and (\ref{eqn:barecoup1}) i.e, 
\begin{equation}
\label{eqn:bareLag}
\begin{split}
&\mathcal{L}_{(b)} = \Bigg[\frac{1}{2} \Big \{(\partial_i B_{(b)})^2+(\partial_t A^i_{(b)})^2 -2 (\partial_t B_{(b)}) (\partial_i A^i_{(b)})\Big\} \\
&\qquad -(\partial_t \phi^*_{(b)})(\partial_t \phi_{(b)}) \Bigg]+m^2_{(b)}  \phi^*_{(b)} \phi_{(b)}-e^2_{(b)} B^2_{(b)} \phi^*_{(b)} \phi_{(b)} \\
&\qquad \qquad- i e_{(b)} B_{(b)} \Big (\phi_{(b)} \partial_t \phi^*_{(b)}-\phi^*_{(b)} \partial_t \phi_{(b)} \Big) 
\end{split}
\end{equation}
This completes the renormalization process for sCED. However, there are several things to note here. First of all, mass and coupling redefinitions have turned out to be heavily gauged. Secondly, the leading divergent terms in the bare mass and bare coupling i.e, (\ref{eqn:renormass}) and (\ref{eqn:renormcoup}) admits a mass shell singularity at $m^2 \to\omega^2$. Off course, one might be tempted to take the limit, $m \to 0$ such that the mass shell singularity term drops. However, this complicates the situation even more as bare mass and bare couplings then become infrared divergent. Recall that the massless limit of sCED is actually a conformal theory at the classical level \cite{Bagchi:2019clu}. The emergence of IR divergences at the quantum level further complicates the matter. An important thing to observe here is that IR divergences are present even in the massless scalar Carrollian theory. For example, consider the Lagrangian for a massless Carrollian $\varphi^4$ theory 
\begin{equation*}
\mathcal{L}= \frac{1}{2} (\partial_t \varphi)^2 -\lambda \varphi^4
\end{equation*}
where $\varphi$ is the scalar field and $\lambda$ is the coupling constant. The propagator $\< \varphi, \varphi\>$ is given by
\begin{equation*}
\< \varphi, \varphi\> =  \frac{i}{\omega^2}
\end{equation*}
It then is obvious that the first order loop correction will require one to evaluate integrals of the type $\sim \bigintsss d\omega  \frac{i}{\omega^2}$, which clearly leads to IR divergences when $\omega \to 0$. The source of these IR divergences is the mass shell singularity and it is a generic feature of the conformal Carrollian theories (presently known). \\[5pt]
 We refrain ourself to expand more on the renormalization such as beta function and renormalization group flow for sCED until the issue of gauge dependence and IR divergences gets settled. Clearly, in this work, we have demonstrated that the standard procedure of renormalizing when applied to Carrollian field theories lead to the violation of conventional arguments of gauge independence of mass and coupling. Further, the bare quantities defined above diverge severely on the mass shell. Lastly, the massless limit renders an IR divergent notion of bare mass and bare coupling which further complicates the renormalization structure. 
Clearly, the renormalization of Carrollian gauge theories is not well understood at the moment and the potential issues mentioned above seem rather unavoidable as of now. Some more work in the Carrollian quantum sector is hereby needed. This paper should thus be viewed as the first step towards exploring the quantum ``properties" of Carrollian field theories.

\section{Conclusion}
\label{section:conclusion}
Let us now summarize our findings. In this paper, we explored the renormalization properties of a massive sCED in $3+1$ dimensions prescribed via functional techniques. We essentially highlighted the potential issues that crop up while renormalizing a Carrollian abelian gauge theory such as sCED (at first order in the perturbation) via standard functional techniques.\\[5pt]
To begin with, we propose an action for massive sCED consistent with Carrollian symmetries. Owing to the symmetries of the action, we construct the associated Noether charges and confirm that the Carrollian algebra is realized at the level of charges. We then implement path integral techniques to explore the renormalization structure of the theory. Since sCED is a gauge theory, we gauge fix the action by implementing the Faddeev-Popov trick. A trivial dimensional analysis suggests that the theory falls into the category of marginally renormalizable theories. We state the Feynman rules for the theory and study the renormalization valid up to the first order in the perturbation. To this end, we evaluate the allowed 1-loop correction to the propagators and the vertices. However, the renormalization condition renders an unphysical notion of mass and coupling, in that they turn out to be gauge dependent. This behaviour bears a stark resemblance to the massive Schwinger model in $1+1$ dimensions where the fermion mass turns out to be gauge dependent. Since mass and coupling strength are physical observables for a theory, the issue of their gauge dependence has to be settled which brings us to the list of open questions that we shall be addressing in our upcoming works. \\[5pt]
The first and most prominent question to address is to have a gauge-independent notion of mass and coupling for a renormalized sCED. Our first guess is to draw on the wisdom from the Lorentzian case. Generally in Lorentz invariant field theories, we employ Nielsen identities to redefine the mass renormalization conditions which then renders us with a gauge-independent notion of renormalized mass. It shall be interesting to see if we can carry out a similar procedure for sCED and establish the gauge independence of mass and coupling. Another possible way is to study the renormalization under quenched rainbow approximation\cite{10.1143/PTP.52.1326}. This approximation has also been used to establish gauge independence of fermion mass for massive Schwinger model \cite{Das:2012qz}\cite{Das:2013vua}\cite{Das:2013iha}. However, one serious limitation of this approach is that higher loop correction becomes computationally difficult making it harder to establish the renormalizability at higher order in the perturbation.\\[5pt] 
A natural question that follows; is gauge dependence (of mass and coupling) a generic feature of all gauge Carrollian quantum field theories? To this end, an interesting thing to study would be to see how the renormalization conditions modify if we replace a massive Carrollian scalar with a massive Carrollian fermion.
One of the research work that we are currently looking forward to is the canonical quantization of Carrollian theories. Some work in this direction is already in progress and shall be reported in the near future. Extending the quantization program to the case of conformal Carrollian theories would be one of the directions of future works.  

\section*{Acknowledgements}
We would like to thank Kinjal Banerjee and Rudranil Basu for careful reading of the manuscript and several useful discussions. We would also like to thank JaxoDraw\cite{Binosi:2003yf} for developing their free Java program for drawing the Feynman diagrams. AM is supported by the Royal Society URF of Jelle Hartong through the Enhanced Research Expenses 2021 Award vide grant number: RF\texttt{\symbol{92}}ERE\texttt{\symbol{92}}210139.

\appendix

\section{Propagators in position space}
\label{section:prop}
The propagators for sCED in \emph{momentum} space (where, $\omega \; \text{and} \; p_i$ are the Fourier transform of \; $\partial_t$\; and $\partial_i$ respectively) are
\begin{eqnarray}
&\big< B, B \big> =&\frac{-i \;\xi}{\omega^2}\\[4pt]
&\big< B, A_i \big>=&\frac{-i \; \xi p_i}{\omega^3}\\[4pt]
&\big< A_i, A_j \big>=&i \frac{\delta_{ij}}{\omega^2}-\frac{i \xi p_i p_j}{\omega^4}\\[4pt]
&\big< \phi^*, \phi \big> =&\frac{i}{-\omega^2+m^2}
\end{eqnarray}
We shall now write down the propagators in the position space. This can be achieved by taking their inverse Fourier transform. In position space, the propagator takes on the following form,
\begin{eqnarray}
&\big< B, B \big> =& i \xi \sqrt{\frac{\pi}{2}} \; t \; \text{sgn}(t)\; \delta^3(r) \\[4pt]
&\big< B, A_i \big> =& i \xi \frac{\pi}{2} \; t^2\; \text{sgn}(t)\; \partial_i \delta^3 (r) \\
&\big< A_i, A_j \big> =& i \xi\delta_{ij} \pi \; t\; \text{sgn}(t)\;\delta^3(r)+i \xi \sqrt{\frac{\pi^3}{18}} \; t^3\; \text{sgn}(t)\; \partial_i \partial_j \delta^3 (r)\\[4pt]
&\big< \phi^*, \phi \big> =& \frac{e^{-imt} (-1+e^{2 i mt}) \pi^{\frac{3}{2}}}{\sqrt{2} m} \text{sgn}(t)\; \delta^3(r) 
\end{eqnarray}
where $sgn(t)$ is the \emph{signum} function for time $t$ and $\delta^3 (r) $ is the Dirac delta function capturing the ultra local behaviour.

\section{Carrollian geometry: A crash course!}
\label{section:geometry}
A Carrollian manifold is defined as a quadruple $(\mathbb{C}, \gamma, \chi, \Gamma)$ known as Carrollian structure where
\begin{eqnarray*}
&\mathbb{C} &\equiv \text{a smooth $d+1$ dimensional manifold}\\
&\chi&\equiv \text{a nowhere vanishing vector field}\\
&\gamma&\equiv \text{a degenerate metric tensor whose kernel $(ker \xi)$ is generated by $\chi$}\\
&\Gamma&\equiv \text{affine connection on $\mathbb{C}$}
\end{eqnarray*}
Note that the degeneracy in the metric $\gamma$ does not allow to define $\Gamma$ uniquely by the pair $(\gamma, \chi)$. The simplest Carroll structure we can think of is the flat Carroll structure which in the coordinate chart $(t,x,y,z)$ is given by
\begin{equation}
\label{eqn:Carr str}
\mathbb{C}= \mathbb{R}^3 \times \mathbb{R} \qquad, \qquad \gamma=\gamma_{ab} dx^a \otimes dx^b \qquad, \qquad \chi= \frac{\partial}{\partial t} \qquad, \qquad \Gamma=0
\end{equation}
where $a,b$ are the space and time indices that runs from $0,i$ and $\gamma_{ab}$ is a degenerate metric i.e,
\begin{equation*}
\gamma_{ab}= \begin{pmatrix}
0&0\\
0& \delta_{ij}
\end{pmatrix}
\end{equation*}
With Carrollian structure at our disposal, we can define Carroll group as the set of diffeomorphism that preserves the metric $\gamma$, the vector field $\chi$ and the affine connection $\Gamma$ ; also known as $\chi$ preserving isometries i.e for the vector field $X \in \mathbb{C}$ we have, 
\begin{eqnarray*}
\label{eqn:sym4}
&\pounds_{\text{\tiny{$X$}}} \gamma_{ab}&=0\\
\label{eqn:sym5}
&\pounds_{\text{\tiny{$X$}}} \chi^a&=0\\
\label{eqn:sym6}
&\pounds_{\text{\tiny{$X$}}} \Gamma&=0
\end{eqnarray*}
For a flat Carrollian structure \eqref{eqn:Carr str}, the $\xi$ preserving isometries takes the following form,
\begin{equation}
\label{eqn:geomXcar}
X=(\omega^i_j x^j+\beta^i)\partial_i+(\alpha-\gamma^i x_i)\partial_t
\end{equation}
where $\omega^i_j \in O(3)$, $\beta^i, \gamma^i \in \mathbb{R}^3$ and $\alpha \in \mathbb{R}$. Reading off the symmetry generators (\ref{eqn:generators}) from (\ref{eqn:geomXcar}) is then pretty much straight forward.

\section{Canonical analysis}
\label{section:dirac}
In this appendix, we are interested in exploring the gauge \emph{nature} of Carrollian electrodynamics (CED). Consider the Lagrangian for CED,
\begin{equation}
\label{eqn:lagCED}
L=\bigintssss d^3x \;\; \frac{1}{2} \Big \{(\partial_i B)^2+(\partial_t A_i)^2 -2 (\partial_t B) (\partial_i A_i)\Big\}
\end{equation}
To understand the gauge structure, we perform Dirac constraint analysis. Our starting point is the canonical Hamiltonian $(H_c)$ of the system i.e, 
\begin{equation}
\label{eqn:canH} 
H_c =\bigintssss d^3x \;\;\frac{1}{2}\Big((\pi^i)^2 -(\partial_i B)^2 \Big)
\end{equation}
where $\pi^i$ is the canonical momentum associated to $A_i$. It should be noted that while working out the Legendre transformation of (\ref{eqn:lagCED}) we encounter the following primary constraint:
\begin{equation}
\label{eqn:cons}
C_1= \pi_B+ \partial_i A_i
\end{equation} 
where $\pi_B$ is the canonical momenta associated to $B$. The admission of the primary constraint in the theory calls for the augmentation of the canonical Hamiltonian with a Lagrange multiplier $(\lambda)$. Following Dirac's notation \cite{dirac2013lectures}, we call the augmented canonical Hamiltonian as the total Hamiltonian $(H_t)$,
\begin{equation}
\label{eqn:ht}
H_t= \bigintssss d^3x\;\; \Big( \frac{1}{2} (\pi^i)^2 -\frac{1}{2} (\partial_i B)^2 +\lambda (\pi_B+\partial_i A_i) \Big)
\end{equation}
The consistency check for $C_1$ leads to the secondary constraint $C_2$ in the theory:
\begin{equation}
\label{eqn:c2}
\{C_1, H_t\}= \partial^2 B+\partial_i \pi^i \; \equiv \; C_2 \approx 0
\end{equation} 
A consistency check for $C_2$ reveals that no further constraints are present in the theory.  A trivial calculation can now be carried out to see that $C_1$ and $C_2$ Poisson commute i.e, $\{C_1, C_2\} =0$, thus making them first class constraint. The existence of first class constraint confirms that CED is a gauge theory. Since there are only two scalar first class constraints, the physical phase space dimension (in $d=3+1$ space time dimension) turns out to be 4, just like we have in the case of Lorentzian QED.  Now to construct an arbitrary gauge generator $G$ we first smear the two first class constraint by arbitrary test functions $\alpha_1$ and $\alpha_2$ i.e, 
\begin{eqnarray}
&\mathcal{C}_1[\alpha_1]&= \bigintssss d^3x\;\; \alpha_1 \Big(\pi_B+ \partial_i A_i\Big)\\
&\mathcal{C}_2[\alpha_2]&= \bigintssss d^3x\;\; \alpha_2 \Big(\partial^2 B+\partial_i \pi^i  \Big)
\end{eqnarray}
The generator of gauge transformation $G$ is defined as a linear combination of $\mathcal{C}_1$ and $\mathcal{C}_2$ such that
\begin{equation}
G= \mathcal{C}_1[\alpha_1]+\mathcal{C}_2[\alpha_2]
\end{equation}
The gauge transformation generated by $G$ on $B$ and $A_i$ can be worked out by the off-shell condition \cite{Banerjee:1999hu}:
\begin{equation}
\label{eqn:psi}
\delta_G \frac{d}{dt} \psi = \frac{d}{dt} \delta_G \psi
\end{equation}
where $\psi$ is any dynamical function and $\delta_G$ is the transformation generated by the gauge generator $G$ via
\begin{equation}
\delta_G F(q,p)= \{F, G\}
\end{equation}
for any phase space function $F$. Choosing $F$ to be $B$ and $A_i$, we can arrive at the following gauge transformation for CED
\begin{eqnarray}
&\delta_G B&= \alpha_1\\
&\delta_G A_i &= -\partial_i \alpha_2
\end{eqnarray}
Note that $\alpha_1$ and $\alpha_2$ can not be independent (as one of the first class constraint is a primary constraint) and are related to each other via $\partial_i (-\alpha_1+\partial_t\alpha_2)=0$.

\section{Renormalization of Carrollian $\varphi^4$ theory}
\label{section:CSP}
\begin{figure}[h]
\centering
\includegraphics[scale=0.4]{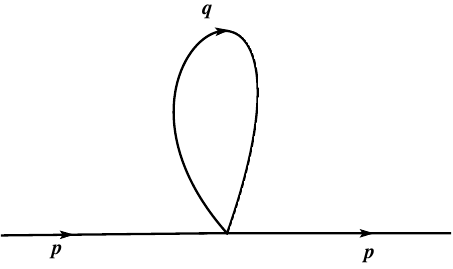}
\caption{Correction to the propagator}
\label{fig:pc}
\end{figure}

\begin{figure}[h]
\centering
\includegraphics[scale=0.4]{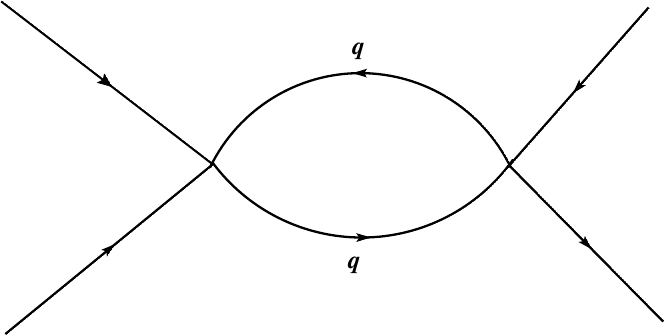}
\caption{Correction to the vertex}
\label{fig:vc}
\end{figure}

Consider the Lagrangian $\mathcal{L}$, for an interacting massive Carrollian scalar field $\varphi$
\begin{equation}
\label{eqn:philag}
\mathcal{L}= \frac{1}{2} (\partial_t \varphi)^2 -\frac{1}{2} m^2 \varphi^2 -\frac{1}{4!} g \varphi^4
\end{equation}
The mass dimensions of the coupling $g$ turns out to be zero and thus the theory is marginally renormalizable. The momentum space Feynman rules for the theory are given in \autoref{fig:feyn}.
\begin{table}
\begin{center}
\begin{tabular} { | m{0.5cm} | m{1.5cm} | m{4cm} | m{1cm} |}
\hline
1. & $\;\; \Big<\varphi, \varphi\Big>$ &\qquad \includegraphics[scale=0.45]{prop1}& $\frac{i}{\omega^2-m^2}$\\[7pt]
\hline
2. & $\quad V_{\tiny{\varphi^4}}$ & \quad \includegraphics[scale=0.45]{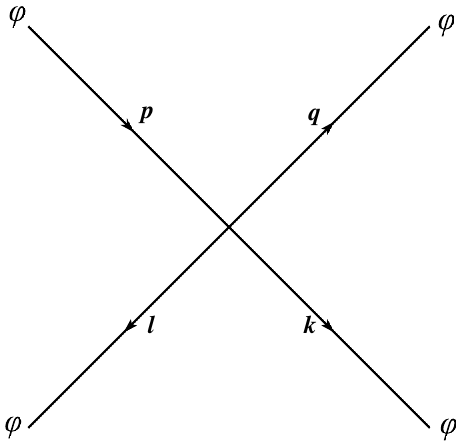}& $-i g$\\
\hline
\end{tabular}
\caption{\label{fig:feyn}Feynman rules}
\end{center}
\end{table}
Owing to the self interaction, there are two possible corrections at 1 loop in the theory viz. the correction to the propagator and correction to the vertex. The Feynman diagram for the propagator correction is given by \autoref{fig:pc}.
The corresponding integral $I_1$ evaluates to
\begin{equation}
I_1= \frac{4 i \pi^2 \Lambda^3}{3 m} g
\end{equation}
where, $\Lambda$ is the UV momentum cut off. Following the renormalization scheme, it can be checked that a mass counter term is required to absorb the divergence in the propagator i.e,\begin{equation}
(\mathcal{L}_{1})_{counter} =-\frac{1}{2} \mu^2 \varphi^2
\end{equation}
where $\mu^2 = \frac{4 \pi^2 \Lambda^3}{3 m} g$. Next, the correction to the vertex (\autoref{fig:vc}) evaluates to 
\begin{equation}
I_2= -\frac{4 \pi^2 \Lambda^3 i}{6 m^3} g^2
\end{equation}
Trivially, it can be checked that the counter term required to absorb the divergence in the vertex is 
\begin{equation}
(\mathcal{L}_{2})_{counter} =-\frac{1}{4!} g C \varphi^4
\end{equation}
where $C=\frac{4 \pi^2 g^2 \Lambda^3}{6 m^3}$. Adding counter terms to (\ref{eqn:philag}), results in the bare Lagrangian $\mathcal{L}_{(b)}$
\begin{equation}
\mathcal{L}_{(b)}=\frac{1}{2} (\partial_t \varphi_{(b)})^2 -\frac{1}{2} m^2_{(b)} \varphi^2_{(b)} -\frac{1}{4!} g_{(b)} \varphi^4_{(b)}
\end{equation}
where $\varphi_{(b)} =\varphi$, $m_{(b)}^2 =m^2+\mu^2$ and $g_{(b)}= g(1+C)$. It is instructive to note here that the theory is renormalizable at 1 loop and does not admit any IR divergences.  Also, unlike the sCED, renormalized mass and coupling are well defined.

\bibliographystyle{unsrt}
\bibliography{Ref}

\end{document}